\documentclass[prd,preprintnumbers,superscriptaddress,showpacs,eqsecnum,twocolumn,nofootinbib]{revtex4}

\usepackage{graphicx}
\usepackage{amssymb}
\usepackage{amsmath}
\usepackage{amsthm}
\usepackage{psfrag}

\def\be{\begin{equation}}
\def\ee{\end{equation}}
\def\ba{\begin{eqnarray}}
\def\ea{\end{eqnarray}}

\def\Kp{\tilde{K}}
\def\hp{\tilde{h}}
\def\Kc{K}
\def\hc{h}
\def\Dp{\tilde{D}}
\def\Dc{D}
\def\Rp{\tilde{\mathcal{R}}}
\def\Rc{\mathcal{R}}
\def\qp{\tilde{q}}
\def\qc{q}

\def\kp{\tilde{k}}
\def\kc{k}

\def\kpt{\tilde{k}^0}
\def\kct{{k}^0}

\def\grad{\nabla}

\theoremstyle{definition}

\newcommand{\ck}{\mathcal{L}}
\newcommand{\ckt}{l}
\usepackage{graphicx}
\usepackage[usenames,dvipsnames]{color}

\begin{document}

\title{On the existence of initial data containing isolated black holes}

\newcommand*{\AEI}{Max-Planck-Institut f\"ur Gravitationsphysik,
  Albert-Einstein-Institut, Am M\"uhlenberg 1, D-14476 Golm, Germany}
\newcommand*{\MEU}{Laboratoire de l'Univers et de ses Th\'eories, UMR 8102
  du C.N.R.S., Observatoire de Paris, F-92195 Meudon Cedex, France}

\author{Sergio Dain}\email{dain@aei.mpg.de}\affiliation{\AEI}
\author{Jos\'e Luis Jaramillo}\email{jose-luis.jaramillo@obspm.fr}\affiliation{\MEU}
\author{Badri Krishnan}\email{badri.krishnan@aei.mpg.de}\affiliation{\AEI}

\date{DRAFT VERSION; $ $Date: 2004/12/13 18:05:54 $ $}

\begin{abstract}

We present a general construction of initial data for
Einstein's equations containing an arbitrary number of black holes,
each of which is instantaneously in equilibrium.  Each black hole is
taken to be a marginally trapped surface and plays the role of the
inner boundary of the Cauchy surface.  The black hole is taken to be
instantaneously isolated if its outgoing null rays are
shear-free. Starting from the choice of a conformal metric and the
freely specifiable part of the extrinsic curvature in the bulk, we
give a prescription for choosing the shape of the inner boundaries and
the boundary conditions that must be imposed there. We show rigorously
that with these choices, the resulting non-linear elliptic system
always admits solutions. 

\end{abstract}

\preprint{AEI-2004-122}
\maketitle
 

\section{Introduction}
\label{sec:intro}

In this paper we consider the problem of specifying quasi-equilibrium 
multi-black hole initial data for the vacuum Einstein equations.
This problem is important as a starting point for the numerical
simulations of binary black hole spacetimes. The initial data should
be such that the black holes are in equilibrium with time independent
horizon geometries.  Furthermore, the entire geometry of the spatial
slice should, in an appropriate sense, be in quasi-equilibrium with
minimal spurious radiation content.  If both these criteria are
satisfied, and if the black holes are in a roughly circular orbit
around each other, we could reasonably expect the calculated
gravitational waveforms to correspond to observations by gravitational
wave detectors.  

The problem of finding such initial data has received a lot of
attention in the numerical relativity literature and significant
progress has been made in the last few years.  The original numerical
work addressing this issue was due to Cook in 2002 \cite{Cook02} 
who, working in the so-called conformal thin sandwich (CTS)
decomposition \cite{York99} of the initial data, proposed a set of
conditions for solving the initial value problem subject to the
quasi-equilibrium conditions.  More recent developments in this
direction can be found in \cite{Cook04,Yo04}.
See Cook \cite{Cook94} and Gourgoulhon {\it et 
  al}. \cite{Gourgoulhon02a,Gourgoulhon02b}  for an earlier numerical 
study of binary black hole initial data which is in quasi-equilibrium
in the bulk.  The case of quasi-equilibrium configurations in presence
of matter is discussed in e.g. \cite{Baumgarte04}.  
See also \cite{Schaefer03} for an approach to this
problem based on the post-Newtonian expansion.   

Independently of the numerical work, on the analytical side, a
quasi-local approach to black hole physics was developed by Ashtekar 
\textit{et al.} (see e.g. \cite{Ashtekar99, Ashtekar00a,Ashtekar00b,
  Ashtekar04}).  This work lead to the notion of an isolated horizon
which models a black hole in equilibrium in an otherwise dynamical
spacetime.  While isolated horizons are defined in the full four
dimensional spacetime, it would seem natural that this framework
should also have 
implications for the construction of quasi-equilibrium initial data.
This issue was studied in some detail by Jaramillo \textit{et al.} 
\cite{Jaramillo04}.  They used the isolated horizon formalism, also
working in the CTS decomposition, to arrive at a set of  
boundary conditions for the constrained parameters of the initial data
which is similar to that obtained by Cook, but with certain additional
constraints on the boundary values of the otherwise free data. While
the initial results are promising in both approaches, it is not yet
settled if they will finally succeed.      

The aim of this work is to point out some potential mathematical
difficulties with the approaches of \cite{Cook02, Jaramillo04, Cook04}
and to suggest a resolution of these difficulties.  Our aim is less
ambitious than \cite{Cook02, Jaramillo04, Cook04} in the sense that we
only consider the issue of the appropriate boundary conditions at the
horizon: i) we do not make any statement about the conditions that
must be satisfied in the bulk to ensure that the entire data set is in
quasi-equilibrium and ii) we do not discuss gauge conditions for the
evolution equations. Because we do not address  i) and ii), we will work
with the standard conformal transverse-traceless (CTT) decomposition
of the initial data \cite{York73} and not with the CTS as in
\cite{Cook02, Jaramillo04, Cook04}, because the former simplify the
discussion. However we expect our main results to be relevant also in
other decompositions including the CTS.

The conceptual issues have to do with the boundary conditions for the
momentum constraint. All the decompositions involve an elliptic
equation for a vector $\beta^a$.  The references \cite{Cook02,
  Jaramillo04, Cook04} all use a Dirichlet condition for $\beta^a$
which adapts the time evolution vector to the properties of the
horizon.  This requirement becomes intertwined with the initial data
construction in the CTS decomposition since the latter can be
interpreted to involve also the notion of time evolution.  However,
strictly speaking, the initial data construction is distinct from the
choice of gauge for time evolution.  More importantly, as already
pointed out in \cite{Jaramillo04}, a Dirichlet condition for $\beta^a$
entails potential problems for the very existence of solutions for the
Hamiltonian constraint. We will show that a new type of geometric
boundary condition for $\beta^a$ can be used, which allows us to
apply the theorems proved by Dain and Maxwell \cite{Dain03,Maxwell03}
in order to get a rigorous proof of existence of solutions for the
resulting non-linear equations under appropriate assumptions.  These
assumptions, which involve restrictions on the shape of the inner
boundaries and some inequalities the free data should satisfy, can be
checked  \emph{a priori}.  The advantage of this analysis
is that the solution is guaranteed to exist in more general cases than
the ones studied numerically so far; in particular, in non
conformally flat cases. This is important for two reason: first, the
conformal geometry of a quasi-equilibrium black hole initial data is
still unknown (though it is clear that it cannot be conformally flat
because the Kerr metric does not admit conformally flat slices
\cite{Valiente04}). Second, one is usually interested not just in a
single solution, but in families of solutions which depend smoothly
on the relevant parameters of the problems, like separation distances, 
individual spins and linear momentum, etc. For example, one notion of
quasi-equilibrium is defined as a variational problem for a family of
solutions (cf. \cite{Cook02} and also \cite{Cook00} and references
therein).   Thus, we would like to ensure that the initial value
equations admit solutions for the widest possible range of
parameters and free data.  

Section \ref{sec:basics} sets up notation and states the
mathematical problem we want to solve.  Section \ref{sec:main} gives
the main result and compares it to earlier work, section
\ref{sec:details} consists of a mathematical proof of the main result,
and section \ref{sec:conclusion} summarizes our results and suggests
directions for future work.

\section{The conformal method and non-expanding horizons}
\label{sec:basics}

The problem we want to solve is to find initial data on a spatial
slice $M$ for the vacuum Einstein's equations.  Thus, we want to find
a 3-metric $\hp_{ab}$ and a second fundamental form 
\begin{equation}
\Kp_{ab} := -\hp_a^c\hp_b^b\grad_c\tau_d = - \frac{1}{2}{\cal L}_\tau \hp_{ab}\,,
\end{equation}
such that the constraints are satisfied:
\begin{eqnarray}
\Dp_a (\Kp^{ab} - \Kp\hp^{ab}) &=& 0 \,,\\
\Rp + \Kp^2 - \Kp_{ab}\Kp^{ab} &=& 0 \,.
\end{eqnarray}
Here $\tau^a$ is the unit timelike normal to $M$, and $\grad_a$ is the
four dimensional covariant derivative.  Our sign convention for
$\Kp_{ab}$ corresponds to what is commonly used in the numerical
relativity literature.  $\Dp_a$ is the derivative operator compatible
with $\hp_{ab}$, and $\Rp$ is its scalar curvature.  We restrict
ourselves to maximal slices, i.e. we always take $\Kp = 0$.  It is
important to note that very little is known about solutions of the
constraint equations (with or without inner boundaries) in the case
when $\Kp$ is not nearly constant (see the recent review
\cite{Bartnik:2002cw} and references therein).

The metric and the traceless part
of the extrinsic curvature are conformally rescaled 
\begin{equation}
\hp_{ab} = \psi^{4}\hc_{ab}\,, \qquad \Kp^{ab} = \psi^{-10}\Kc^{ab}\,.
\end{equation}
As a rule, physical tensors on $M$ will be denoted with a tilde to
distinguish them from the corresponding conformally rescaled
quantities (note the opposite convention with respect to Refs.
\cite{Cook02,Cook04,Jaramillo04}).  In terms of the conformally
rescaled quantities $(\hc_{ab},\Kc_{ab})$, the constraint equations
become
\begin{eqnarray}
\Dc_a\Kc^{ab} &=& 0, \label{eq:momconstr} \\
L_{\hc} \psi &=& -\frac{1}{8}\Kc_{ab}\Kc^{ab}\psi^{-7}, \label{eq:hamconstr}
\end{eqnarray}
where $L_{\hc}$ is the conformally invariant Laplacian operator:
$L_{\hc} = \Delta - \Rc /8$; $\Delta := D_aD^a$ is the Laplacian,
$D_a$ is the derivative 
operator compatible with $\hc_{ab}$, and $\Rc$ is its scalar
curvature.  To solve the momentum constraint, we decompose (the  
traceless) $\Kc_{ab}$ according to York's prescription \cite{York73}:
\begin{equation} \label{eq:decomp}
\Kc_{ab} = (\mathcal{L}\beta)_{ab} - Q_{ab},
\end{equation}
where $\beta^a$ is a vector field on $M$, $\mathcal{L}$ is the
conformal Killing operator defined as 
\begin{equation}
(\mathcal{L}\beta)_{ab} \equiv 2\Dc_{(a}\beta_{b)} -
    \frac{2}{3}\hc_{ab}\Dc_c\beta^c,
\end{equation}
and $Q_{ab}$ is a freely specifiable symmetric and traceless tensor.
The decomposition of $\Kc_{ab}$ given by equation (\ref{eq:decomp}) is
known as the conformal transverse-traceless decomposition (CTT) but it
is not the only possibility.  In fact, the currently more commonly used 
decomposition in numerical relativity is the so called conformal
thin-sandwich (CTS) decomposition also proposed originally by York
\cite{York99}.    

Using the CTT decomposition, the momentum constraint
(\ref{eq:momconstr}) becomes  
\begin{equation} \label{eq:betaelliptic}
\Delta_L\beta^a = J^a\,.
\end{equation}
Here we have defined 
\begin{equation}
\label{eq:delta}
\Delta_L\beta^a := D_b(\mathcal{L}\beta)^{ab} = \Delta\beta^a +
\frac{1}{3}\Dc^a\Dc_b\beta^b + \Rc_b^a\beta^b
\end{equation}
and $J^a := \Dc_bQ^{ab}$.  Thus, we have to solve the system of
elliptic equations (\ref{eq:hamconstr}) and (\ref{eq:betaelliptic}) on
a domain $M\subset\mathbb{R}^3$ subject to certain boundary conditions
which will occupy us for the rest of this paper.  
\begin{figure}
  \begin{center}
  \psfrag{na}{$n^a$}
  \psfrag{la}{$\ell^a$}
  \psfrag{st}{$\tilde{s}^a$}
  \psfrag{S}{$S$}
  \psfrag{ta}{$\tau^a$}
  \psfrag{M}{$(M,\hp_{ab},\Kp_{ab})$}
  \includegraphics[height=3.7cm]{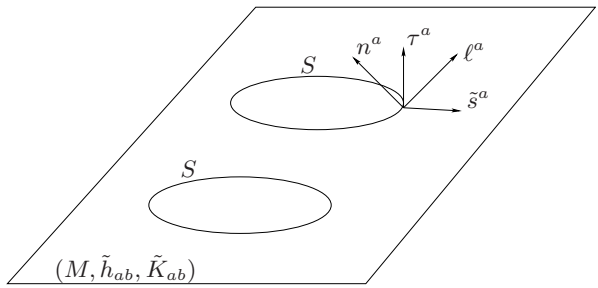}
  \caption{The inner boundary $S$ of the Cauchy surface
  $M$.  In this figure the inner boundary consists of two disconnected
  spherical surfaces.  }\label{fig:id}  
  \end{center}
\end{figure}

We wish to solve equations (\ref{eq:hamconstr}) and
(\ref{eq:betaelliptic}) on a domain $M$ which is $\mathbb{R}^3$ with
an arbitrary finite number of compact sets excised from it.  The inner
boundary will be denoted by $S$, and in general, it is allowed to have
many disconnected components. The excision surface $S$ serves as an
inner boundary for our problem and it represents the surface of the
black hole.  This is understood to mean that each connected component
of $S$ is an outer marginally trapped surface, i.e. when it is viewed
as a closed two surface in the full four-dimensional spacetime
manifold, the congruence of future directed outgoing null geodesics
orthogonal to it has zero expansion.  This is depicted in figure
\ref{fig:id} which shows the Cauchy surface $M$ with inner boundary
$S$.  The unit spacelike normal to $S$ (with respect to the physical
metric $\hp_{ab}$ and pointing towards spatial infinity) is
$\tilde{s}^a$, the unit timelike normal to $M$ is $\tau^a$, and 
\begin{equation}\label{eq:outgoing}  
\ell^a := \tau^a + \tilde{s}^a
\end{equation}
is a fiducial outgoing null normal to $S$.  The ingoing null normal to
$S$ is  
\begin{equation} \label{eq:ingoing}
n^a = \tau^a - \tilde{s}^a\,.
\end{equation}
The conformally rescaled normal $s^a$ is defined as
\begin{equation}
\tilde{s}^a = \psi^{-2}s^a\,.
\end{equation}
The induced physical 2-metric on $S$ is denoted by $\qp_{ab}$ and it
is conformally rescaled as
\begin{equation}
\qp_{ab} = \psi^4\qc_{ab} \,.
\end{equation}
The projection operator onto $S$ is $\qp_a^b = \qc_a^b$.  
The second fundamental form of the inner boundary is defined by
\begin{equation} 
 \label{eq:kp} 
\kp_{ab}= \qp_a^c\qp_b^d\Dp_c\tilde s_d.
\end{equation}
We denote by $\kp$ the trace and by $\kpt_{ab}$ the trace-free part of
$\kp_{ab}$, that is
\begin{equation} 
 \label{eq:kpt} 
\kp_{ab}= \kpt_{ab}+ \frac{1}{2}\kp \qp_{ab}, \quad \kp=\qp^{ab}\kp_{ab}. 
\end{equation} 

With this
notation, we can write down the expression for the expansion $\Theta_+$
of $\ell^a$: 
\begin{eqnarray}
\Theta_+ &=& \tilde{q}^{ab}\grad_a\ell_b  \nonumber \\
&=& \Kp_{ab} \tilde{s}^a\tilde{s}^b - \Kp  +\kp.
\end{eqnarray}

In terms of the conformally rescaled fields and using again $\Kp=0$, 
this is equivalent to 
\begin{equation} \label{eq:expansion}
\Theta_+ = \psi^{-3}\left(4s^a\Dc_a\psi + \psi\kc  + \psi^{-3}\Kc_{ab}s^as^b\right) \,.
\end{equation}
Our first requirement on $S$ is that the expansion $\Theta_+$ must
vanish. 
This is also known as the apparent horizon boundary condition in the
literature.  The problem of constructing initial data satisfying this
boundary condition has already been studied \cite{Dain03,Maxwell03}.
This boundary condition does not place any restrictions on the
physical parameters of the black hole.  We wish to impose the
condition that the black hole is isolated in a certain specific
sense. 

Our notion of $S$ being isolated is based on the isolated horizon
framework \cite{Ashtekar99,Ashtekar00a,Ashtekar00b,Dreyer02}.  This is a
quasi-local framework to study black holes without relying on the
global notion of an event horizon; see \cite{Ashtekar04} for a general
review.  It has found many applications in black hole mechanics,
mathematical relativity, quantum gravity, and also in numerical
relativity \cite{Dreyer02,Jaramillo04}.  An isolated horizon can be
viewed as the world tube $\mathcal{T}$ of marginally trapped surfaces
obtained by time evolution, in the case when $\mathcal{T}$ is null.  
In this case, it can be shown that the area of the black
holes is constant in time and also the flux of gravitational waves
falling into the black hole vanishes
\cite{Hayward93,Ashtekar02,Ashtekar03}. In the general case,
$\mathcal{T}$ is expected to be spacelike, thereby representing a
dynamical black hole.  This is described by a dynamical horizon
\cite{Ashtekar02,Ashtekar03}. Trapping horizons
\cite{Hayward93,Hayward04} can describe the null and spacelike cases
in a unified framework.  

For our purposes, we do not need the details regarding isolated
horizons.  We only need to know that in vacuum, the condition for
the world tube being null is equivalent to the vanishing of the
\emph{shear} $\sigma_{ab}$ of $\ell^a$ \cite{Hayward93,Dreyer02}.
This is essentially a consequence of the Raychaudhuri equation whose
expression, for a future directed geodesic $\ell^a$ with affine
parameter $\lambda$, is
\begin{equation}
  \label{eq:Raych}
\frac{d\Theta_+}{d\lambda}= -\frac{1}{2}\Theta_+ - \sigma_{ab}\sigma^{ab}
+ \omega_{ab}\omega^{ab} - R_{cd}\ell^c\ell^d \,.
\end{equation}
Here $\omega_{ab}$ is the twist of $\ell^a$ and $R_{ab}$ is the
spacetime Ricci tensor; note that $R_{ab}\ell^a\ell^b=0$ in vacuum.  
If $\Theta_+$ is initially zero, and taking the null rays to be
surface forming ($\omega_{ab}=0$), then a non-zero shear would imply 
that $\Theta_+$ will be non-zero at a later time.  Thus the apparent
horizon cannot evolve along $\ell^a$, and the black hole will not be
isolated; see \cite{Hayward93,Ashtekar03} for further  
details and discussion.  Conversely, if we take $S$ to evolve strictly
along $\ell^a$ so that $\mathcal{T}$ is null and $\Theta_+$ is zero at
all times, we get the 
condition that the shear vanishes identically, which in turn can be
shown to imply that the rate of change of the area of $S$ is zero, and
$S$ can be considered to be isolated.  In the presence of matter, we
must impose, say, the null energy condition to get analogous results.
The condition  $\sigma_{ab}=0$ has also been considered independently
by Cook \cite{Cook02}.  In the language of the isolated
horizon framework, this condition would guarantee that $S$ is a
cross-section of an infinitesimal \emph{non-expanding horizon}
\cite{Ashtekar00a}.

The shear of $\ell^a$ is defined as the tracefree part of the
projection of $\grad_{(a}\ell_{b)}$ onto $S$:
\begin{equation}
\sigma_{ab} = \qp_a^c\qp_b^d\grad_{(c}\ell_{d)} - \frac{1}{2}\Theta_+
\qp_{ab} \,.
\end{equation}
In terms of the physical fields it is given by
\begin{equation}
  \label{eq:1}
  \sigma_{ab}=-\tilde q^c_a \tilde q^d_b \tilde K_{cd} +
  \frac{1}{2} \tilde
  q_{ab} \tilde q^{cd}\tilde K_{cd} + \kpt_{ab}\,,
\end{equation}
and, in terms of the conformally rescaled fields, the shear is written as
\begin{equation}
\label{eq:shear}
\sigma_{ab} = \psi^{-2}\left(-q_a^cq_b^d\Kc_{cd} +
\frac{1}{2}\qc_{ab}\qc^{cd}\Kc_{cd}\right) + \psi^2\kct_{ab} \,.
\end{equation}
Here $k_{ab}$ is the extrinsic curvature of $S$ embedded in $M$ with
respect to the conformal metric $\hc_{ab}$
\begin{equation}
k_{ab} = \qc_a^c\qc_b^d\Dc_c s_d,
\end{equation}
and, as before, 
\begin{equation} 
 \label{eq:kpc} 
\kc_{ab}= \kct_{ab}+ \frac{1}{2}\kc \qc_{ab}, \quad \kc=\qc^{ab}\kc_{ab}. 
\end{equation} 

Our task is to prescribe appropriate boundary conditions for
$(\psi,\beta^a)$ so that equations (\ref{eq:hamconstr}) and
(\ref{eq:betaelliptic}) have a regular solution with $\psi$ everywhere
positive, and $\Theta_+=0$, $\sigma_{ab}=0$ at $S$.

\section{The main result}
\label{sec:main}

\subsection{The sign of $K_{ab}s^as^b$}
\label{subsec:kabsasb}

Let us start with the condition $\Theta_+=0$ which, from equation
(\ref{eq:expansion}), can be written as
\begin{equation} \label{eq:psibc}
4s^a\Dc_a\psi  = - \psi\kc -\psi^{-3}\Kc_{ab}s^as^b\,.
\end{equation}
We would like to use this as the boundary condition for solving
equation (\ref{eq:hamconstr}) for $\psi$, together with the standard
condition at infinity
\begin{equation}
  \label{eq:binfty}
  \lim_{r\to \infty} \psi =1.
\end{equation}
Let us first discuss some general properties of Eqs.
(\ref{eq:hamconstr}), \eqref{eq:psibc} and \eqref{eq:binfty}.

We begin with the physical meaning of the sign of $\Kc_{ab}s^as^b$. In
a realistic collapse, the boundary is expected to not only satisfy
$\Theta_+=0$, but also to be a future marginally trapped surface, that
is $\Theta_-\leq 0$, where $\Theta_-$ is the expansion of the ingoing
null-normal $n^a$ defined in Eq. (\ref{eq:ingoing}). For example, a
surface with  
$\Theta_+=0$ and $\Theta_- >0$ is not expected to be present in a
realistic gravitational collapse of matter; these surfaces are located
on the inner null boundary of the left quadrant of the Kruskal diagram
(region IV of Fig. 6.9 in \cite{Wald84}). The expansion $\Theta_-$ is
given by
\begin{align}
\Theta_- & = -\Kp + \Kp_{ab} \tilde{s}^a\tilde{s}^b -\kp\\
 &= \psi^{-3}\left(-4s^a\Dc_a\psi - \psi\kc  +
   \psi^{-3}\Kc_{ab}s^as^b\right) ,\label{eq:inexpansion}
\end{align}
where the second line applies if the data is maximal ($\Kp=0$). 

Under this assumption, we get that $\Theta_+ + \Theta_-=
2\Kp_{ab} \tilde{s}^a\tilde{s}^b$, then for a future marginally
trapped surface on a maximal slice we always have 
\begin{equation} 
 \label{eq:Kss} 
\Kc_{ab}s^as^b\leq 0.
\end{equation}
Also, as mentioned below, it turns out that we also need to control
the sign of $K_{ab}s^as^b$ in order to provide necessary conditions
for the existence of solutions of Eqs. (\ref{eq:hamconstr}),
\eqref{eq:psibc} and \eqref{eq:binfty}. 

On the other hand, if
we were to impose purely Dirichlet boundary conditions on $\beta^a$,
then we could not guarantee that $K_{ab}s^as^b$ has a definite sign at the
boundary.  This is simply because equation (\ref{eq:decomp}) implies
that $K_{ab}s^as^b$ involves  radial derivatives of $\beta^a$ which
cannot be controlled by Dirichlet conditions:
\begin{equation}
K_{ab}s^as^b = (\mathcal{L}\beta)_{ab} s^as^b- Q_{ab}s^as^b\,.
\end{equation}
This conclusion is also  valid for  the conformal
thin-sandwich decomposition.  This is because also in the CTS
decomposition, the extrinsic curvature is written in terms of the
derivative of $\beta^a$ and $\Kp_{ab}\tilde{s}^a\tilde{s}^b$ will
again involve radial derivatives of $\beta^a$. 

We would like to prescribe $(\mathcal{L}\beta)_{ab}s^as^b$ as free
data on $S$ and, at the same time, be able to enforce $\sigma_{ab}=0$
on $S$. The main result of this paper is that this is indeed possible.

\subsection{Existence of solutions to the Hamiltonian constraint}
\label{subsec:hamconstr}

Let us now review the condition telling us about the existence of
solutions to equation (\ref{eq:hamconstr}) with boundary conditions
\eqref{eq:psibc} and \eqref{eq:binfty}.  The conformal factor $\psi$
appears in the denominator of both equations (\ref{eq:psibc}) and
\eqref{eq:hamconstr}, hence these equations are singular if the
conformal factor vanishes, which is in agreement with the fact that
the physical metric $\hp_{ab}$ does not make sense if the conformal
factor is zero at some point. For general conformal metrics and
boundaries, these equations will have no solutions. To illustrate
this, let us consider the time symmetric case ($K_{ab}=0$).  For the
case with no inner boundaries, necessary and sufficient conditions on
the conformal metric $\hc_{ab}$ for the solvability of Eq.
(\ref{eq:hamconstr}) and \eqref{eq:binfty} has been studied in
\cite{Cantor81} and \cite{Maxwell03}. This condition can be written
in term of the following quantity known as the Yamabe
invariant\footnote{In this article we always assume the dimension of
  $M$ is $3$, however all of the following discussion can be
  generalized to arbitrary dimensions, see \cite{Maxwell03}.}  
\begin{equation} 
 \label{eq:yamabe} 
\lambda_h= \inf_{\varphi\in C^\infty_c(M), \, \varphi \not \equiv0}
\frac{\int_M (8D_a\varphi 
D^a\varphi+ \Rc\varphi^2) }{|| \varphi||^2_{L^6}},
\end{equation}
where $C^\infty_c(M)$ denotes functions with compact support in $M$,
$||\varphi||^2_{L^6}=(\int_M |\varphi|^6 )^{1/6}$, and $\varphi
\not\equiv 0$ means that $\varphi$ cannot be identically zero
everywhere. The number $\lambda_h$ is conformally invariant in the
following sense: let $\theta$ be any smooth and positive function, and
let $\hat h_{ab}=\theta h_{ab}$; then $\lambda_{\hat h}=\lambda_h$.
The Yamabe invariant has a long history. It was discovered by Yamabe
for compact 
manifolds and later studied by many authors; see, for example, the
review \cite{Lee87}.

A solution of (\ref{eq:hamconstr})--\eqref{eq:binfty} (with
$K_{ab}=0$) exists if and
only if $\lambda_h > 0$.  Every metric with $\Rc\geq 0$ (the flat
metric, for example) satisfies this condition.  It is obvious from
\eqref{eq:yamabe} that $\Rc\geq 0$ implies $\lambda_h \geq 0$,
however, to prove that in fact $\lambda_h > 0$ is non
trivial. However,  there
exist  metrics which do not satisfy $\lambda_h > 0$ (see
\cite{Cantor81} for an explicit, axisymmetric example). For
these conformal metrics there are  no solutions of
\eqref{eq:hamconstr}--\eqref{eq:binfty}.  

If we include inner boundaries, there exists a generalization of this
condition (cf. \cite{Maxwell03}) in terms of the following conformal
invariant which is a generalization of the Yamabe invariant:
\begin{equation} 
 \label{eq:yamabe2} 
\lambda_{h,S}= \inf_{\varphi\in C^\infty_c(M), \, \varphi\not\equiv
  0}\frac{\int_M (8D_a\varphi 
D^a\varphi+ \Rc\varphi^2) -2\oint_S \kc \varphi^2 }{|| \varphi||^2_{L^6}}.
\end{equation}
This invariant has been studied also for the compact case in
\cite{Escobar92}. We have the following result proved in
\cite{Maxwell03}. 
A solution of Eqs. (\ref{eq:hamconstr}), \eqref{eq:psibc} and
\eqref{eq:binfty}, with $K_{ab}=0$, exists if and only if
$\lambda_{h,S}> 0$.  Note that now $\lambda_{h,S}$ depends on the
choice of the boundary. In particular, note that the boundary term in
\eqref{eq:yamabe2} has a minus sign.\footnote{Unfortunately, there
  exists in the literature different conventions for the signs of
  $K_{ab}$, $k$, and $s^a$. In this article we have used what seems to
  be the standard conventions in numerical relativity. The relation of
  our present convention with the ones in \cite{Maxwell03} and
  \cite{Dain03} is the following. The second fundamental form $K_{ab}$
  is denoted by $\sigma_{ab}$ in \cite{Maxwell03}, and let us denote
  by $\bar K_{ab}$ the one in \cite{Dain03}. Then we have
  $K_{ab}=\sigma_{ab}=-\bar K_{ab}$. The mean curvature of the
  boundary (in our notation $k$) is denoted by $h$ in \cite{Maxwell03}
  and by $H$ in \cite{Dain03}. We have $k=-2h=-H$. The normals are
  denoted by $\nu^a$ in both \cite{Maxwell03} and \cite{Dain03} (they
  use the same choice for it). We have $s^a=-\nu^a$.}  It can be
proved that any metric with $\Rc\geq 0$ and boundary with $k\leq 0$
have $\lambda_{h,S}>0$ \cite{Maxwell03}. As before, in this case it follows
directly from the definition that $\lambda_{h,S}\geq 0$,
but to prove that it is strictly positive requires extra work. Note that
this does not apply to the flat metric with spheres as boundaries,
since for spheres of radius $r_0$ we have $k=2/r_0 > 0$. However, the
flat metric with an arbitrary number of non-intersecting spheres as
inner boundaries satisfies $\lambda_{h,S}>0$, provided they are not
too close to each other. This can be seen as follows. Assume that we
have only one sphere of radius $r_0$ centered at the origin. Take the 
conformal factor $\theta=1+r_0/r$, and define the rescaled metric
$\hat h_{ab}= \theta^4\delta_{ab}$ (where $\delta_{ab}$ is the flat
metric). This is, of course, the Schwarzschild initial data. This
metric satisfies $\hat R =0$ and the boundary $r=r_0$
is a minimal surface with respect of the metric $\hat h_{ab}$, i.e,
$\hat k =0$. From the previous discussion it then follows that
$\lambda_{\hat h, S}>0$, and since it is conformally invariant we have
$\lambda_{\delta, S}=\lambda_{\hat h, S} >0$. In order to generalize
for more spheres we will use the well known Misner initial data:
Misner shows in Lemma 3 of \cite{Misner63} that there exists a
conformal factor $\theta$ such that $\hat h_{ab}=\theta^4
\delta_{ab}$, $\hat R=0$, with the boundaries of the spheres being
minimal surfaces with respect to the metric $\hat h_{ab}$, i.e, $\hat
k =0$. Then we have also in this case $\lambda_{\delta,
  S}=\lambda_{\hat h, S} >0$. For two spheres, the condition that the
spheres are not too close used in this Lemma, just means that they
do not touch each other. 

It is easy to construct more general examples of metrics which satisfy
$\lambda_{h,S} > 0$. As pointed out in \cite{Maxwell03}, if we take
any metric on $\mathbb{R}^3$ which satisfies $\Rc \geq 0$ and we excise
appropriate small spheres on it, we get 
$\lambda_{h,S} > 0$ on $M$ (see \cite{Maxwell03} for details).

For the non-time symmetric case, if we assume the maximal condition
$K=0$ and no inner boundaries, then is clear that every maximal
initial data satisfies $\lambda_{h} > 0$ because
$\Rp= \tilde K_{ab}\tilde K^{ab}\geq 0$. This is also a sufficient
condition (with appropriate fall-off conditions for $K_{ab}$ at
infinity) for the conformal metric to ensure the existence of
solutions to the non linear equation (\ref{eq:hamconstr}) with
boundary conditions \eqref{eq:binfty} (see \cite{Bartnik:2002cw},
\cite{Maxwell04} and references therein). 

If we include inner boundaries, still requiring the condition $K=0$,
it is not 
clear if the condition $\lambda_{h,S}>0$ is necessary. However, it has
been proved in \cite{Dain03, Maxwell03} that it is a \emph{sufficient}
condition together with additional assumptions on $k$ and
$K_{ab}s^as^b$. Moreover in these references the condition
$\lambda_{h,S}>0$ plays an essential role.  Even if the condition
$\lambda_{h,S}>0$ turns out not to be necessary in generic situations,
it is very likely that it will still be relevant for the black hole
problem.  The reason being that for these data, time symmetry arises
as a limit when the linear and angular momentum of the black holes are
zero, and in this case, as discussed above, $\lambda_{h,S}>0$ is a
necessary and sufficient condition.   

The assumptions on $k$ and $K_{ab}s^as^b$ are always
made in some representative metric in the class $\lambda_{h,S}>0$.
That is, these assumptions are not conformally invariant. The two main
examples are the following. 
If we assume that the conformal metric satisfies $\Rc\geq 0$ and 
 $k\leq 0$ (this automatically implies $\lambda_{h,S}>0$) and we
 assume $K_{ab}s^as^b\geq 0$ at the boundary $S$, then there is always a
solution of Eqs.  (\ref{eq:hamconstr}), \eqref{eq:psibc} and
\eqref{eq:binfty} (with $K=0$ and appropriate fall-off conditions
for $K_{ab}$). This case can be obtained from \cite{Dain03} making a
time reversion (that is, $t^a \rightarrow -t^a$, $\Theta_+ \rightarrow
- \Theta_-$, $\Theta_- \rightarrow - \Theta_+$, $K_{ab}\to -K_{ab}$).
This example seems to be physically relevant only if
$K_{ab}s^as^b=0$ at the boundary since, if this is not the case,
then $\Theta_- > 0$. The second example was studied in
\cite{Maxwell03}. The conformal metric is assumed to satisfy
$\lambda_{h,S}>0$ and, in addition, $\Rc=0$, $k>0$ (the flat metric with
spheres as boundaries satisfies these conditions). It is assumed also that
$-k\leq K_{ab}s^as^b\leq 0$. In this case we get future trapped surfaces. 

It is important to note that for an arbitrary metric which satisfies
$\lambda_{h,S}>0$, it is possible to calculate (solving a linear
equation) a conformal factor $\theta$ such that the rescaled metric
$\hat h_{ab} =\theta^4 h_{ab}$ satisfies $\hat R=0$ and $\hat k>0$ or
(with, of course, a different $\theta$) $\hat R\geq 0$ and $\hat k\leq
0$ (cf. \cite{Maxwell03}).

We have seen that in both cases the existence theorems assume a
definite sign for $K_{ab}s^as^b$ at the boundary. This is essential in
order to control the positivity of the conformal factor using the
boundary condition \eqref{eq:psibc} and the maximum principle. In
order to be able to control $K_{ab}s^as^b$ at the boundary, it is
necessary to use a boundary condition for the momentum constraint
\eqref{eq:momconstr} that allows us to prescribe $K_{ab}s^as^b$ as free
data. In \cite{Dain03, Maxwell03} this has been achieved using the
following Neumann type boundary condition for $\beta^a$ on $S$:
\begin{equation}
  \label{eq:8}
 (\mathcal{L}\beta)_{ab} s^a=0\,.  
\end{equation}
However, we will see in the next section that with this boundary
condition, we cannot enforce $\sigma_{ab}=0$ at the boundary. In order
to do this a new kind of boundary condition will be needed.

\subsection{Boundary conditions for the momentum constraint} 
\label{subsec:momconstr}

Let us now focus on the condition $\sigma_{ab}=0$ with $\sigma_{ab}$
given by equation (\ref{eq:shear}). First note that the two terms in
\eqref{eq:shear} involve different powers of the conformal factor
$\psi$. Since $\psi$ is the unknown solution of the nonlinear equation
\eqref{eq:hamconstr}, it will be very difficult to get $\sigma_{ab}=0$
without requiring that both these terms vanish independently.  In
particular, this requirement will imply that the shear of the ingoing null
geodesics along $n^a$ is also zero. Then, the second term of
(\ref{eq:shear}) suggests that we impose  
$\kct_{ab}=0$. This is a condition on the shape of the boundary $S$,
and is also known as the umbilical condition. For example a sphere in
the flat metric satisfies this. Note that this condition is
conformally invariant, that is $\kct_{ab}=0 \iff \kpt_{ab}=0$.

Having made the above choice for $k_{ab}$, we are then left with the
first term of (\ref{eq:shear}). Before discussing the general case, let
us mention some important examples of initial data which not only
satisfy $\sigma_{ab}=0$ at $S$, but also have $\kp_{ab}=0$.    
 
The first example  is provided by the Misner solution, already
discussed above.  The fact that these data satisfy $\sigma_{ab}=0$ at
$S$ can be directly verified from \eqref{eq:1}: due to the reflection
isometry of these data we automatically have $\kp_{ab}=0$
\cite{Gibbons72}, and $\Kp_{ab}=0$ because the data is time 
symmetric. 

The second important example is the Kerr initial data for the
Boyer-Lindquist slicing. Since Kerr is stationary, it is clear that
$\sigma_{ab}=0$. Moreover, these data also have an isometry which
leaves the horizon fixed, then we also have  $\kp_{ab}=0$ in this
case. Note that although the Kerr data are not time symmetric, the
terms with $\Kp_{ab}$ in \eqref{eq:1} still vanish.   

Finally, some of the Bowen-York data \cite{Bowen80} also satisfy these
conditions. These are the data for one black hole with spin (defined
by Eq. (10) in \cite{Bowen80}) and for the so called ``negative
inversion'' single black hole case with linear momentum (defined by
$K^{-}_{ij}$ in Eq. 9 of \cite{Bowen80}). For these cases, we have an
isometry which leaves the horizon fixed, so that
$\kp_{ab}=0$. Moreover, one can explicitly check that the first term
in (\ref{eq:shear}) vanishes for these conformal second fundamental
forms.  

Let us now return to the general case. We make a decomposition of
$\beta$ into its normal and tangential parts with respect to $S$ 
\begin{equation}
  \label{eq:2}
  \beta^a=bs^a + \beta^a_{||},
\end{equation}
where $\beta^a_{||}s_a=0$, $b=\beta^as_a$. 
We insert  \eqref{eq:decomp} in \eqref{eq:shear}, we use the
decomposition \eqref{eq:2} and after some computations we find
\begin{multline}
  \label{eq:3}
 \sigma_{ab}= (\psi^2-2\psi^{-2}b) \kct_{ab} \\
+\psi^{-2}\left( \qc_a^c\qc_b^d Q_{cd} -
\frac{1}{2}\qc_{ab}\qc^{cd}Q_{cd} -(\ckt\beta_{||})_{ab}  \right)\,.
\end{multline}
Here $\ckt$ is the conformal Killing operator on the surface $S$,
that is 
\begin{equation}
  \label{eq:4}
   (\ckt m)_{ab}= d_{(a}m_{b)}-\qc_{ab}d_cm^c,
\end{equation}
where $d$ is the covariant derivative with respect to $q_{ab}$,
$m^a$ is any tangential vector to $S$ ($s^am_a=0$) and
$m_a=q_{ab}m^b$. 

The important point is that Eq. \eqref{eq:3}, if we assume
$\kct_{ab}=0$, only contains tangential derivatives of
$\beta_{||}$\footnote{Incidentally, this is also the case in 
\cite{Jaramillo04,Cook04}, where it is achieved by canceling the
factor multiplying $k^0_{ab}$ by means of a Dirichlet boundary
condition on $b$. The price is the loss of the control on
$K_{ab}s^as^b$.}.  These derivatives can be controlled using Dirichlet    
boundary conditions for  $\beta_{||}$.  Thus, if we choose
$(\ckt\beta_{||})_{ab}$ and $Q_{ab}$ appropriately, we can ensure  
the vanishing of the shear at $S$.  It turns out that this is
in fact a valid boundary condition for $\beta^a$.  

The complete set of boundary conditions on $\beta^a$ for solving
equation (\ref{eq:betaelliptic}) is  
\begin{align}
s^as^b(\mathcal{L}\beta)_{ab} &= f,\label{eq:6}\\
\beta^a_{||} & =\varphi^a\,,\label{eq:7}
\end{align}
where $f$ and $\varphi^a$ are to be prescribed a priori, subject to
certain restrictions mentioned below.

With these choices, the problem is solved as follows. Chose any
$\varphi^a$  and let $Q_{ab}$ be such that 
\begin{equation} \label{eq:5}
(\ckt\varphi)_{ab} - \qc_a^c\qc_b^d Q_{cd} +
\frac{1}{2}\qc_{ab}\qc^{cd}Q_{cd}=0, 
\end{equation}
(alternatively, one could prescribe $Q_{ab}$  and attempt to 
solve for $\varphi^a$).
Solve equation (\ref{eq:betaelliptic}) for $\beta^a$ using the
boundary conditions \eqref{eq:6}-~\eqref{eq:7}. Then (provided
$\kct_{ab}=0$) we will have $\sigma_{ab}=0$.  Furthermore,
$K_{ab}s^as^b=f-Q_{ab}s^as^b$ is a free data which can be chosen to be
non-positive by an appropriate choice of the quantities $Q_{ab}s^as^b$
and $f$ which have, up to now, not been constrained at all.  The proof
that it is always possible to solve equation (\ref{eq:betaelliptic})
with boundary conditions \eqref{eq:6}-~\eqref{eq:7} is given in
section \ref{sec:details}.  

Having solved the momentum constraint and having retained
$K_{ab}s^as^b$ as a free data, we can now solve the Hamiltonian
constraint (\ref{eq:hamconstr}) for the conformal factor subject to
the boundary conditions (\ref{eq:psibc}) and (\ref{eq:binfty}).  The
possible choices of this free data have been given earlier in section
\ref{subsec:momconstr}.  Recall that to solve the momentum constraint,
we only have fixed the tracefree part $k_{ab}^0$; the trace $k$
is still free.  The physically most relevant case 
is  when we want to obtain future trapped surfaces. As discussed
 in section
\ref{subsec:momconstr} the two possibilities are
\begin{enumerate}
\item[i)]$\Rc\geq 0$, $k\leq 0$
and $K_{ab}s^as^b=0$.  Note that all the examples discussed above
satisfy $K_{ab}s^as^b=0$.  This condition implies 
$\Theta_+=\Theta_-=0$ on $S$.
\item[ii)] The more general choice
 $\Rc=0$, $k>0$, and $K_{ab}s^ss^b$ such that  
\begin{equation}\label{eq:signk}
-k \leq K_{ab}s^as^b \leq 0\,.
\end{equation}
\end{enumerate}
In ii) the case $K_{ab}s^as^b=0$ is also included. However, in
practice, for non conformally flat metrics, it is perhaps more
convenient to work with i) because in this case the conditions
$\Rc\geq 0$, $k\leq 0$ involve only inequalities which are easier to
achieve (for example, by a small perturbation of a given metric) than
the condition $\Rc=0$. 

Just like the Dirichlet and Neumann conditions, the boundary
conditions \eqref{eq:6}--\eqref{eq:7} are natural 
for the operator $\Delta_L$.  This is easily seen by considering the  
Green formula for the non flat operator $\Delta_L$:
\begin{equation}
\label{eq:green1}
  \int_\Omega(\ck\beta)^{ab}(\ck\xi)_{ab}= -\int_\Omega \beta^a \Delta_L
\xi_a + \oint_{\partial \Omega}  (\ck \beta)_{ab}s^a \xi^b\,.
\end{equation}
Taking $\xi^a$ to be a solution of the homogeneous problem:
\begin{equation}
\Delta_L\xi^a = 0\,,
\end{equation}
with boundary conditions
\begin{eqnarray}
s^as^b(\mathcal{L}\xi)_{ab} &=& 0\,, \\
\xi^a_{||} &=& 0\,,
\end{eqnarray}
and $\beta^a = \xi^a$, the boundary
term in equation (\ref{eq:green1}) can be seen to vanish. Thus we see
immediately that $\xi^a$ must satisfy $(\mathcal{L}\xi)_{ab} = 0$. 
Therefore, as for the Dirichlet and Neumann (i.e. equation
(\ref{eq:8})) cases, the kernel of the
homogeneous problem with our new boundary conditions consists 
only of conformal Killing vectors.  
The only difference is that in the Dirichlet case the kernel is empty
because there are no conformal Killing vectors which vanish at the
boundary. In the Newmann case every conformal Killing vector is in
the kernel because $(\mathcal{L}\xi)_{ab}s^b$ vanishes identically.  
For our present conditions, the only conformal Killing
vectors that are in the kernel are the ones which are normal to the
boundary.   This suggests that these boundary conditions are well posed, and we shall
show in section \ref{sec:details} that this is indeed so.  

Finally, we want to point out that the new boundary conditions
\eqref{eq:6}--~\eqref{eq:7} can be used in both references
\cite{Dain03, Maxwell03} in replacement of \eqref{eq:8}, all the
results presented there will hold without any essential change. 

\subsection{The 2+1 decomposition}
\label{subsec:2plus1}
While the above description is a complete specification of the
problem, in order to get a better feeling of the nature of this
elliptic system, we decompose the 
elliptic system into its radial and tangential parts, and show that
the above boundary conditions translate into a Robin-type condition
for the function $b$ appearing in equation (\ref{eq:2}).  
Using the 2+1 decomposition (\ref{eq:2}) of the shift to express
the boundary conditions (\ref{eq:6}) and (\ref{eq:7}) on the compact 
excised surface $S$, we find
\begin{align}
2 s^a D_a b - k b &= \frac{3}{2} f + d_a\varphi^a - 2 \varphi^a
d_a \ln N \,,\label{eq:bc_b}\\
\beta^a_{||} &=\varphi^a\,.\label{eq:bc_V}
\end{align}
Here $S$ is characterized as the inverse image of a constant value $r_o$ 
by the height function $r$, and $N$ is the normalizing factor of $s^a$ 
such that $s_a= N D_a r$, and $s^a s_a=1$. 

This decomposition suggests a general 2+1 splitting of $M$, provided
by the slicing defined by $r$, as a procedure to solve the elliptical equation 
(\ref{eq:betaelliptic}) for the shift. Extending (\ref{eq:2}) to the whole
of $M$, we find for $\beta^a_{||}$:
\begin{equation}
\Delta_L \beta^a_{||} = J^a - \Delta_L(b s^a) \equiv J^a_{||} \label{eq:V} \ .
\end{equation}
We can enforce the {\it radial} part of $J^a_{||}$ to vanish,
if we impose the following elliptic equation on $b$
\begin{equation}
 J^a_{||}s_a = 0 \Leftrightarrow \Delta_L(b s^a) s_a = J^a s_a \equiv J_\perp \ .
\end{equation}
Expanding the differential operator we find
\begin{align}
J_\perp &=D_a D^a b + \frac{1}{3} s^a s^c D_a D_ c b \label{eq:b} \\
&+ \frac{1}{3} D_a b 
\left( k s^a - d^a \ln N \right )   \nonumber \\
&+b \left(\Rc _{ac} s^as^c + \frac{1}{3} D^a k - k_{ac} k^{ac} 
- (d_a \ln N) (d^a \ln N)\right).  \nonumber
\end{align}
Note that the principal part of the operator defined by \eqref{eq:b}
is not the Laplacian, but it is nevertheless elliptic. 
In this scheme, we calculate the scalar source $J_\perp$ from the
original source $J^a$ by contracting with the normal vector $s^a$.
Then we solve Eq. (\ref{eq:b}) for $b$ by imposing boundary condition
(\ref{eq:bc_b}). With the resulting $b$ we calculate the source
$J^a_{||}$ and solve Eq.  (\ref{eq:V}) for $\beta^a_{||}$ by using the
Dirichlet boundary condition (\ref{eq:bc_V}).

We conclude with some remarks. First, there is a remarkable choice for
$\varphi^a$, namely to prescribe it as a conformal Killing vector on
$S$ (this is condition (49) in \cite{Cook04} or constraint (57) in
\cite{Jaramillo04}), that is $ (\ckt \varphi)_{ab}=0$. This is  a
particular case of condition \eqref{eq:5}. However, this choice seems
to play no fundamental role in our analysis. 

Secondly, it is interesting to remark on a generalization of the above
procedure. In \cite{Dain03}, the free data is chosen in such a way
that it is possible to control the size of both $|\Theta_-|$ and
$|\Theta_+|$.  In other words, it is possible to control, in 
some sense, how trapped the boundary is. The same is possible here
with the shear. This question is of interest if we want to control the
amount of radiation falling into the black hole 
\cite{Hayward93,Ashtekar02,Ashtekar03}.  In equation (\ref{eq:5}), if
we choose $Q_{ab}$ such that the right hand 
side is some $\Sigma_{ab}$ instead of zero, then the elliptic
equations for $\beta^a$ and $\psi$ can still be solved and the final
solution will have an inner boundary with shear
$\sigma_{ab}=-\psi^{-2}\Sigma_{ab}$.  Since $\psi \geq 1$ we get
$|\sigma_{ab}|\leq |\Sigma_{ab}|$. It is also possible to get a lower
bound for $|\sigma_{ab}|$ using the upper bound for $\psi$  obtained
in the existence theorem in \cite{Dain03}.

\section{Elliptic boundary conditions for the momentum constraint}
\label{sec:details}

It is well known that the operator $\Delta_L$ defined by Eq.
\eqref{eq:delta} is elliptic. For a given elliptic operator, a set of
boundary conditions are called elliptic if they satisfy the
Lopatinski-Schapiro conditions, also known as the covering conditions.
For the definition of these conditions as well as the other concepts
used in this section see, for example, the review \cite{Dain04} and
references therein. These conditions are important because an elliptic
operator with elliptic boundary conditions will always have solutions
provided the sources and the boundary values satisfy a finite number
of conditions. 

The operator $\Delta_L$ is (for a three dimensional manifold) of
degree $3$, and we therefore need to prescribe three equations as 
boundary conditions. Let $\Omega$ be a bounded domain in
$\mathbb{R}^3$. The most important example of an elliptic boundary
condition for $\Delta_L$ is the Dirichlet one
\begin{equation}
\label{eq:diri} 
\beta^a=\varphi^a \text{ on } \partial \Omega, 
\end{equation} 
where $\varphi^a$ is an arbitrary vector on the boundary. This condition has
been extensively used in numerical relativity. The analog to the Neumann
condition for $\Delta_L$ is given by 
\begin{equation}
 \label{eq:neumann}
(\ck\beta)_{ab}s^b=\varphi_a \text{ on } \partial \Omega,
\end{equation} 
where $s^a$ is the normal to $\partial \Omega$. 
These conditions are  elliptic.    

In this section we want to prove that the following boundary
conditions for $\Delta_L$ are also elliptic
\begin{align}
s^as^b(\mathcal{L}\beta)_{ab} &=\varphi_1\label{eq:b1}\\
\beta_a m_1^a &=\varphi_2,\label{eq:b2}\\
 \beta_a m_2^a &=\varphi_3,\label{eq:b3}
\end{align}
where $m_1^a$ and $m_2^a$ are tangential and linearly independent vectors at
the boundary.  Note that Eqs.
\eqref{eq:b1}--\eqref{eq:b3} are three linear conditions for $\beta^a$
at the boundary.  In order to write the Lopatinski-Schapiro conditions
we need to define the principal part of the both the operator and the
boundary 
conditions at an arbitrary point $x_0$ on the boundary. For the
operator $\Delta_L$, the principal part is given by the standard
definition. That is, it is given by the terms which contains only two
derivatives. If we choose coordinates at $x_0$ such that
$h_{ab}(x_0)=\delta_{ab}$, then the principal part of $\Delta_L$ is
given by
\begin{equation} 
 \label{eq:pdelta} 
\Delta^0_L= \Delta^0\beta^a +
\frac{1}{3}\partial^a\partial_b\beta^b,
\end{equation}
where $\Delta^0$ is the flat Laplacian and $\partial$ denotes partial
derivatives. 

For the boundary conditions we need to be careful in the definition of the
principal part. If we choose only the terms which contain the highest order
derivatives, then only Eq. \eqref{eq:b1} will survive and this will lead to an
ill posed problem. In order to take into account the fact that terms
of mixed order appear in the boundary operator,
we need to use the general definition of the principal part given by
\cite{Agmon64}. This definition involves integer weights that, with
the notation of \cite{Dain04} and for the operator $\Delta_L$, are
given by $t_\nu=s_\nu=1$ ($\nu=1,2,3$).  For the 
boundary condition, again in the notation of \cite{Dain04}, we have
$r_0=0$, $r_1=r_2=-1$, where $r_0$ corresponds to Eq. \eqref{eq:b1},
$r_1$ to Eq. \eqref{eq:b2}, and $r_2$ to Eq. \eqref{eq:b3}. With this
choice the principal part will include all the equations
\eqref{eq:b1}--\eqref{eq:b3}. That is, the principal part at $x_0$ is
given by   
\begin{equation}
s^as^b(\mathcal{L}^0\beta)_{ab}, \quad
\beta_a m_1^a, \quad  \beta_a m_2^a,
\end{equation}
where
\begin{equation} 
 \label{eq:flatck} 
(\mathcal{L}^0\beta)_{ab} = 2\partial_{(a}\beta_{b)} -
    \frac{2}{3}\delta_{ab}\partial_c\beta^c,
\end{equation} 
and the tangential vectors $m$ are evaluated at $x_0$. 

\begin{figure}
  \begin{center}
  \psfrag{x0}{$x_0$}
  \psfrag{x2}{$(x_1,x_2)$}
  \psfrag{bd}{$\partial\Omega$}
  \psfrag{sa}{$s^a$}
  \psfrag{x3}{$x_3$}
  \psfrag{hp}{$x_3 < 0$} 
  \includegraphics[height=8cm]{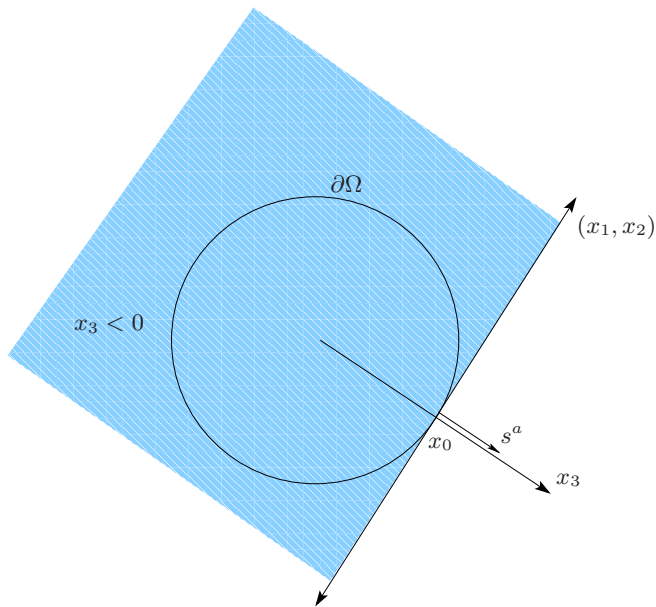}
  \caption{The set-up of the Lopatinski-Schapiro condition.  $x_0$ is 
  a point on the boundary $\partial\Omega$, $(x_1,x_2)$ span the
  tangent plane to $\partial\Omega$ at $x_0$, and $x_3$ is orthogonal
  to $\partial\Omega$; $s^a$ is the unit normal to $\delta\Omega$. The
  shaded region is $x_3 < 0$. }\label{fig:lopatinski}    
  \end{center}
\end{figure}

For a given point $x_0$ at the boundary, we choose coordinates
$(x_1,x_2,x_3)$ such that the normal is given by $s=\partial/\partial
x_3$, and $(x_1,x_2)$ are coordinates on the tangential plane at
$x_0$; see figure \ref{fig:lopatinski}.  
Consider the homogeneous constant coefficient problem, on the half
plane $x_3<0$ with boundary $x_3=0$
\begin{align}
\Delta^0_L\beta^a =0\label{eq:lpo}\\
s^as^b(\mathcal{L}^0\beta)_{ab}=0 \label{eq:lsb1}\\
\beta_a m_1^a=0 \label{eq:lsb2}\\
 \beta_a m_2^a=0 \label{eq:lsb3}.
\end{align}
The boundary conditions are said to satisfy the Lopatinski-Schapiro
conditions if there are no nontrivial solution of
\eqref{eq:lpo}--\eqref{eq:lsb3} of the form 
\begin{equation} 
 \label{eq:exp} 
\beta^j=v^j(x_3)e^{i(\xi_1x_1+\xi_2x_2)} ,
\end{equation}
where $\xi_1$ and $\xi_2$ are arbitrary real numbers, and $v^i(x_3)$
tends to zero exponentially as $x_3\to -\infty$. 
To prove this, the key will be the following Green equation valid
for any $\xi^a$:
\begin{equation}
\label{eq:fgreen1}
  \int_\Omega(\ck^0\beta)^{ab}(\ck^0\xi)_{ab}= -\int_\Omega \beta^a \Delta^0_L
\xi_a + \oint_{\partial \Omega}  (\ck^0 \beta)_{ab}s^a \xi^b\,.
\end{equation}
The proof is very similar to the one given in example 10 of \cite{Dain04} for
the boundary conditions \eqref{eq:diri} and \eqref{eq:neumann}. Let us assume
that there exists a solution of the form \eqref{eq:exp} to equations
\eqref{eq:lpo}--\eqref{eq:lsb3} in the half plane $x_3\leq 0$.   
Let $L_1=2\pi/\xi_1$ and
$L_2=2\pi/\xi_2$. Consider the following  sub-domain contained in the
half plane $x_3\leq 0$: the infinite cubic region $x_3\leq 0$,
$0\leq x_1 \leq L_1$, $0\leq x_2 \leq L_2$.  For this sub-domain we use
equation \eqref{eq:fgreen1} for $\beta^a=\xi^a$. We want to prove that, on
this domain, the boundary integral in \eqref{eq:fgreen1} vanishes for a
solution of the form \eqref{eq:exp}. This is clear for the face 
$x_3= -\infty$ because the solution,
by hypothesis, decay exponentially at infinity. 
Take the face $x_3=0$.  On this face the normal $s^a$ is also the normal to
$\Omega$.  Eqs. \eqref{eq:lsb2}--\eqref{eq:lsb3} 
imply that $\beta^a=\alpha s^a$ on this face for some function
$\alpha$. We use Eq. \eqref{eq:lsb1} to conclude that the integrand in
the  boundary integral vanishes on this face.  On the other faces, the
integrand does not vanish. However, because of 
the choice of $L_1$ and $L_2$, we have that the integrand of opposite
faces are identical. Then, the sum of the boundary integrals vanishes
because the normal is always outwards.  We conclude that
$(\ck^0\beta)_{ab}=0$. But there are no conformal Killing vectors in
flat space which decay to zero at infinity. Hence the
Lopatinski-Schapiro conditions are satisfied, and the boundary
conditions are therefore elliptic.

From the previous discussion, using standard results in elliptic theory, we
deduce that a solution $\beta^a$ of the boundary value problem 
\begin{align}
\Delta_L\beta^a &= J^a \text{ on } \Omega \\
s^as^b(\mathcal{L}\beta)_{ab} &=f \text{ on } \partial \Omega \\
\beta_a m_1^a &=\varphi_1,\text{ on } \partial \Omega \\
\beta_a m_2^a &=\varphi_2, \text{ on } \partial \Omega.
\end{align}
exists if and only if 
\begin{equation} 
 \label{eq:rest} 
\oint_{\partial\Omega}b f= \int_\Omega J^a\xi_a
\end{equation}
for all conformal Killing vectors $\xi^a$ of the metric $h_{ab}$ which
are normal to the boundary of $\Omega$, and where $b$ is defined by
$\xi^a=bs^a$. This can be shown by considering the Green equation
(\ref{eq:green1}) from which we deduce
\begin{eqnarray}
 && \int_\Omega \left(\xi^a\Delta_L \beta_a-  \beta^a \Delta_L
 \xi_a\right) \nonumber \\ &=&  \oint_{\partial \Omega}
 \left((\ck \beta)_{ab}s^a\xi^b - (\ck \xi)_{ab}s^a \beta^b\right)
 \,. \label{eq:green2}
 \end{eqnarray} 
In this equation, set $(\mathcal{L}\xi)_{ab}=0$ (which implies
$\Delta_L\xi^a=0$) and $\xi^a=bs^a$ at the boundary, to immediately
get Eq. (\ref{eq:rest}).   

If the metric admits no
conformal Killing vectors, then there are no restrictions on $J^a$. An
example of a metric with conformal Killing vectors is the flat metric.
If the boundary is a sphere centered at the origin, then $\xi^a=x^a$
is a conformal Killing vector which is normal to the boundary (here we
have assumed coordinates such that $h_{ab}=\delta_{ab}$).

In our case, $J^a=D_bQ^{ab}$ for some trace-free tensor $Q^{ab}$. Then
condition \ref{eq:rest} can be written as
\begin{equation} 
 \label{eq:rest2} 
\oint_{\partial\Omega} \left( b f-  Q_{ab}s^b\xi^a\right) =0.
\end{equation}
for the exterior region $M$ discussed in this article, the
boundary integral \eqref{eq:rest2} contains two terms, one is the
inner boundary $S$ and the other is an integral over the sphere at
infinity.

For simplicity we have considered in this section only bounded
domains, since the new part here is given by the boundary conditions
on the inner boundary.  Conditions at infinity (i.e., fall-off) for
$\beta^a$ are the standard ones, that is $\beta^a\to 0$ at infinity.
See, for example \cite{Maxwell03} where weighted spaces have been used
and \cite{Dain03} where a compactification of the exterior region has
been employed. With this fall-off condition, the kernel is always trivial
because there are no conformal Killing vectors which decays to zero at
infinity. However, in the presence of a conformal Killing vector
normal to the boundary, equation \eqref{eq:rest2} still plays a
role. In this case it relates some asymptotic components of the
solution with a boundary integral; this is the analog of the
restrictions studied in \cite{Beig96, Dain99, Dain03} for the momentum
constraint.

\section{Conclusions} 
\label{sec:conclusion}

In this paper, we have explained why a Dirichlet condition on
$\beta^a$ may be potentially problematic for solving the Hamiltonian
constraint. This is essentially because with a Dirichlet condition it is
not possible to control the sign of $K_{ab}s^as^b$ at the boundary,
since this function depends on \emph{normal} derivatives of $\beta^a$.
To control the sign of this function is important for two reason: the
first one is that for physically interesting solutions, i.e. ones
which contain marginally future trapped surfaces, on maximal slices
$K_{ab}s^as^b$ is always non-positive. The second one, is that with a
definite sign of $K_{ab}s^as^b$ it is possible to prescribe a priori
conditions that guarantee that the solutions of the non linear
equations will always exist. It seems to be very hard to obtain such
conditions without controlling the sign of this function. We have shown
that this can be achieved imposing a Neumann (oblique) boundary
condition on the radial part of $\beta^a$ and Dirichlet conditions on
its tangential parts.  Using the theory of elliptic systems, we have
shown that these boundary conditions are well posed.  

\section*{Acknowledgements}
We are grateful to Abhay Ashtekar for valuable discussions and for
suggesting this problem to us, and to Helmut Friedrich for illuminating
discussions regarding the invariant \eqref{eq:yamabe}.  We are also
grateful to Marcus Ansorg, Eric Gourgoulhon, Francois Limousin, and
Guillermo Mena Marug\'an for valuable discussions.  We also
acknowledge the support of the Albert Einstein Institute and of the
Observatoire de Paris.  SD acknowledges support from the
Sonderforschungsbereich SFB/TR 7 of the Deutsche
Forschungsgemeinschaft. JLJ acknowledges the support of a Marie
Curie Intra-European Fellowship within the 6th European Community
Framework Programme, and the hospitality of the Albert Einstein
Institute.

\bibliography{biblio}

\end{document}